# Astro2020 Science White Paper

# Polarization in Disks

**Thematic Areas:** ☒ Planetary Systems  ☒ Star and Planet Formation
☐ Formation and Evolution of Compact Objects  ☐ Cosmology and Fundamental Physics
☐ Stars and Stellar Evolution  ☐ Resolved Stellar Populations and their Environments
☐ Galaxy Evolution  ☐ Multi-Messenger Astronomy and Astrophysics

**Principal Author:**
Name: Ian Stephens
Institution: CfA/SAO
Email: ian.stephens@cfa.harvard.edu
Phone: 617-496-2197

**Co-authors:** (names and institutions)
Zhi-Yun Li (University of Virginia), Haifeng Yang (Tsinghua University), Akimasa Kataoka (NAOJ), Leslie Looney (University of Illinois), Charles Hull (NAOJ), Manuel Fernández-López (Instituto Argentino de Radioastronomía), Sarah Sadavoy (CfA/SAO), Woojin Kwon (KASI/UST), Satoshi Ohashi (RIKEN), Ryo Tazaki (Tohoku University), Dan Li (NOAO), Thiem Hoang (KASI/UST), Gesa H.-M. Bertrang (MPIA), Carlos Carrasco-González (IRyA-UNAM), William Dent (ALMA/Chile), Satoko Takahashi (JAO, NAOJ), Francesca Bacciotti (INAF-OAA), Felipe O. Alves (MPE), Josep M. Girart (CSIC/IEEC), Qizhou Zhang (CfA/SAO), Ramprasad Rao (ASIAA), Adriana Pohl (MPIA), Marco Padovani (INAF-OAA), Daniele Galli (INAF-OAA), Chin-Fei Lee (ASIAA), Dominique Segura-Cox (MPE)

**Abstract** (optional):

Polarized dust emission outside of disks reveal the magnetic field morphology of molecular clouds. Within disks, however, polarized dust emission can arise from very different mechanisms (e.g., self-scattering), and each of them are useful for constraining physical properties in the disk. For example, these mechanisms allow us to constrain the disk grain size distributions and grain/disk geometries, independent from current methods of measuring these parameters. To accurately model these features and disentangle the various polarization mechanisms, multi-wavelength observations at very high resolution and sensitivity are required. With significant upgrades to current interferometric facilities, we can understand how grains evolve in disks during the planet formation process.

Polarized dust emission in the interstellar medium is widely accepted to trace the direction of magnetic fields, given that the short axis of dust grains is expected to align parallel to the field direction via radiative torques (B-RAT alignment; e.g., Andersson et al. 2015). Since disks are magnetically active and can possibly drive angular momentum transport via magnetorotational instability and disk winds (e.g., Balbus & Hawley 1998; Konigl & Pudritz 2000), observers have wanted to use polarized observations to trace the disk magnetic field morphology for a long time. However, such observations are challenging since they require both spatially resolving the disk (size scales of about 100 au) and reaching high sensitivities. The fraction of the intensity that is polarized is often <1% (Hughes et al. 2009; 2013), i.e., >100 times more sensitivity is needed than a pure continuum study. Before ALMA, polarization in disks was only resolved in the candidate Class 0 disk IRAS 16293–2422B (Rao et al. 2014), the Class I/II disk HL Tau (Stephens et al. 2014), the Class 0 disk L1527 (Segura-Cox et al. 2015), and the candidate high-mass disk Cepheus A HW2 (Fernández-López et al. 2016). However, these observations only resolved polarization in the disk over a few beams, and a much higher resolution and sensitivity (sub)millimeter telescope (e.g., ALMA) was needed to detect polarization in disks.

Recent disk polarimetric observations from ALMA found polarization morphologies that were not consistent with that expected from B-RAT alignment. Around 870 μm, polarization morphologies tend to be uniformly parallel to the minor axis of the disk (e.g., Stephens et al. 2017; Bacciotti et al. 2018; Cox et al. 2018; Hull et al. 2018; Dent et al. 2019). At 3 mm, polarization may change drastically to an azimuthal morphology (Kataoka et al. 2017), and at 1.3 mm the morphology can be a mix of those at 870 μm and 3 mm (Stephens et al. 2017). Figure 1 shows how these morphologies change with wavelength. However, at all wavelengths B-RAT alignment is expected to produce morphologies that are either radial (e.g., Yang et al. 2016a) or radial+azimuthal (Matsakos et al. 2016), which is not observed.

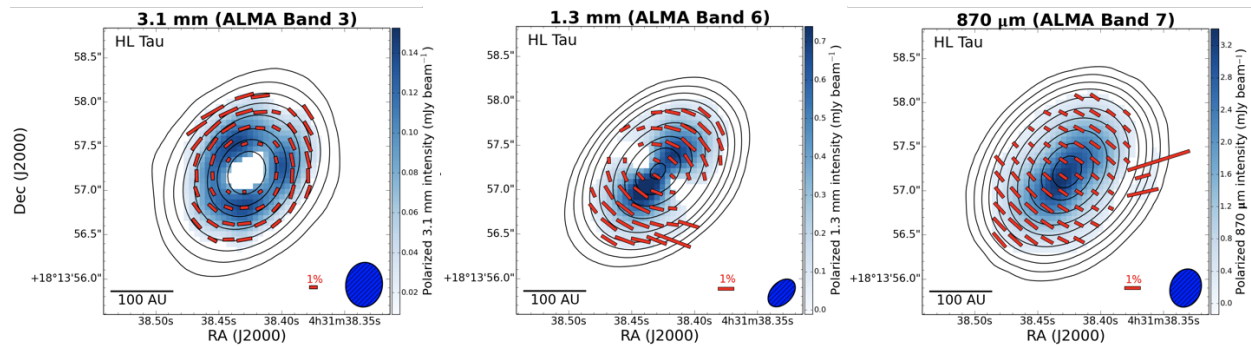

**Figure 1:** Polarization pattern changing with wavelength for the HL Tau disk (Stephens et al. 2017). The color scales show polarized intensity. The polarization morphology is azimuthal at 3.1 mm and uniform at 870 μm. At 1.3 mm, the morphology appears to be a mix of the two.

Based on models and observations, studies have concluded that polarization from disks is generally due to polarization mechanisms other than B-RAT alignment, such as polarization due to dust grain self-scattering. These alternative mechanisms, discussed in more detail later in this white paper, depend on the grain size distribution and the physical properties within the disk. They open a new field in understanding disk grain properties. As such, dust polarization gives insight on grain growth/evolution during the formation of planets. ALMA has just started the revolution of creating a theoretical and observational understanding of polarized dust emission from disks. Nevertheless, to continue making progress, high resolution, multi-wavelength observations are needed.

**Dust Scattering**

Models have shown that the uniform polarization pattern seen from most disks at 870 µm (e.g., right panel of Figure 1) is very likely to come from self-scattering of dust emission (e.g., Kataoka et al. 2015; 2016a, Pohl et al. 2016; Yang et al. 2016a,b, 2017). Scattering is a powerful tool for probing grain sizes in disks, as the polarized intensity of scattering is proportional to the square of the grain size ($a^2$). Figure 2 shows how polarization is expected to change with wavelength for grain distributions with different maximum sizes. Modeling the scattering polarization spectrum also gives insight on grain properties (e.g., the shape, albedo, and porosity; Kataoka et al. 2015). The thickness of the dust layer in the disk can also be inferred. Yang et al. (2017) showed that geometrically thick, inclined disks (e.g., a flared disk) will be more polarized on the disk side that is nearer to the observer and less polarized on the side farther from the observer. Such asymmetric feature has been observed toward several disks (Girart et al. 2018; Bacciotti et al. 2018; Harris et al. 2018). If the disk shows polarization symmetry along the minor axis, then the disk is expected to be geometrically thin (e.g., Stephens et al. 2017, Hull et al. 2018).

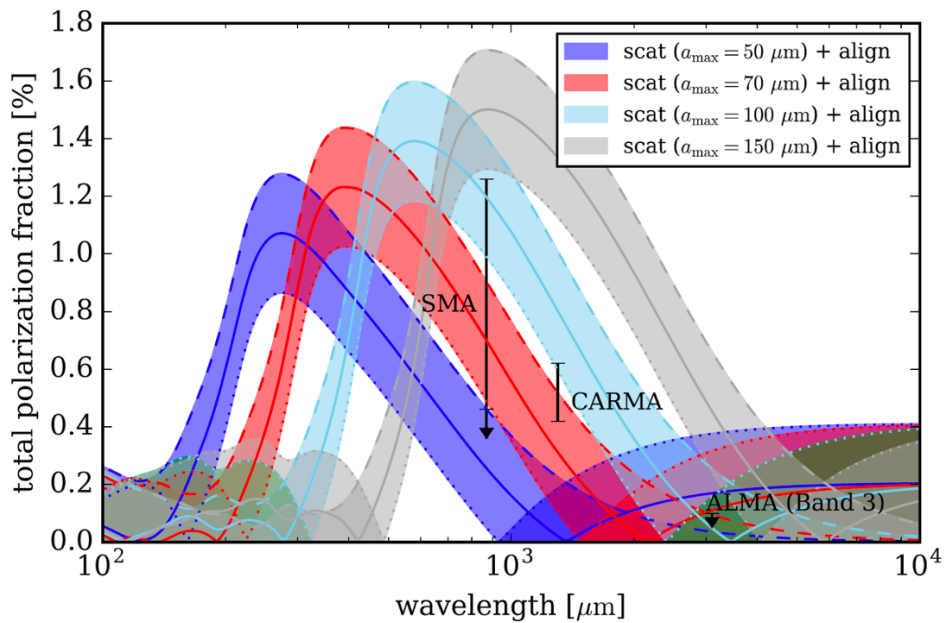

**Figure 2:** The expected polarization spectrum due to self-scattering, (Kataoka et al. 2017). The different colors show different cutoffs for the maximum grain size for the grain distribution. All curves follow a power-law size MRN distribution of n(a) ∝ $a^{-3.5}$. Initial measurements of polarization from HL Tau are shown in black (SMA, CARMA, and ALMA).

The power of these measurements is that they provide an estimate of the grain size that is independent of the dust opacity index, β. Estimates thus far have suggested that the maximum grain sizes in most disks are no more than about 150 µm (e.g., Kataoka et al. 2016b, 2017; Hull et al. 2018). This is in stark contrast to many studies based on β that suggest the maximum grain sizes are larger than a millimeter (e.g., Kwon et al. 2011; Perez et al. 2015; Tazzari et al. 2016; Liu et al. 2017). The dust scattering model must be tuned via multi-wavelength observations that fill the entire polarization spectrum. No study yet has found where the polarization "peak" lies (likely at wavelengths shorter than 870 μm; Figure 2), which is necessary to constrain the scattering model. Moreover, long wavelength observations, such as those greater than 2 mm, are very difficult with current facilities, and can only be observed for a few disks (e.g., Liu et al. 2018).

Grain estimates via measurements of both the dust opacity and the polarization spectrum are also affected by disk substructure, which appears to consist of optically thick rings and optical thin gaps at wavelengths around 1 mm (e.g., Andrews 2018). Very high resolution (few au) polarimetric measurements, capable of detecting polarization fractions to about 0.5%, are needed to understand the effects of substructure on the polarization measurements made with coarser observations. These high resolution observations will also probe differences in grain sizes for different parts of a disk's substructure.

**Polarization at Long Wavelengths**

Figure 1 shows an azimuthal polarization morphology at 3 mm for HL Tau, which is clearly different than the morphology observed at 870 μm. Other disks also show similar azimuthal morphologies at 3 mm (Harrison et al. submitted). This azimuthal morphology was initially thought to be due to "k-RAT alignment," a mechanism where grains align with their long axis perpendicular to the radiation field. For k-RAT alignment, an azimuthal morphology is expected. For large grains (e.g., millimeter-sized), k-RAT alignment is more likely to align grains than B-RAT (Lazarian & Hoang 2007; Tazaki et al. 2017). Mechanical alignment via the Gold (1952) mechanism, where grains align with their long axis in the streaming direction, is also expected to produce an azimuthal morphology, though Gold alignment is expected to be ineffective for subsonic flows. Yang et al. (2019) modeled the expected morphology for both k-RAT and Gold alignment. While both mechanisms can reasonably fit the polarization morphology, neither can produce the observed polarization intensity pattern (see Figure 3). Mechanical torques due to interaction of gas flow with irregular grains can also align dust grains with the long axis perpendicular to the magnetic field or gas flow direction (Lazarian & Hoang 2007; Hoang et al. 2018), but would also produce an azimuthal variation that is not observed (Yang et al. 2019).

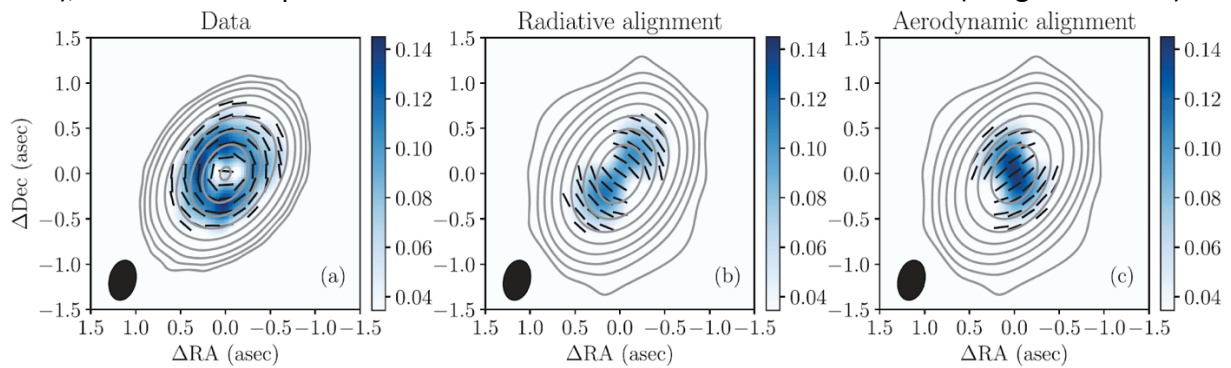

**Figure 3:** Polarimetric maps from Yang et al. (2019). Color scales show the polarized intensity. Left: Polarization pattern observed for HL Tau at 3.1 mm. Middle and Right: Modeled polarization patterns expected for k-RAT (radiative) alignment and Gold (aerodynamic) alignment, respectively. Neither model can reproduce HL Tau's polarized intensity morphology.

Currently, there is no clear explanation for what causes polarization at 3 mm for HL Tau and other disks. This lack of explanation implies a fundamental piece is missing in our understanding of dust physics within disks. Creating a cohesive theoretical model is difficult since very few long wavelength (>2 mm) observations have been made. Moreover, some polarization morphologies at long wavelengths are not completely azimuthal (e.g., Alves et al. 2018; Liu et al. 2019; Harrison et al. submitted). Part of the complication is due to varying optical depths at these wavelengths. Since disks become less bright at longer wavelengths, observations are very time intensive. Polarimetric observations with the VLA toward disks without envelopes have yet to result in a detection due to the lack of sensitivity. The SKA will measure disk polarization at very long wavelengths (≳2 cm), but only can do so for the brightest disks.

**Magnetic Fields**

As previously mentioned, the initial goal of disk polarimetric observations was to measure the magnetic field in circumstellar disks. Magnetic field morphologies from polarization may be difficult if not impossible to map in some disks. If grains are large, k-RAT is preferred over B-RAT to align the grains. If grains are comparable to the wavelength and the optical depths are at least moderate ($\tau \gtrsim 0.5$), self-scattering is likely to dominate polarization. Nevertheless, there have been multiple studies suggesting that magnetic fields are causing grain alignment in at least some parts of the disk (Yang et al. 2016a; Alves et al. 2018; Lee et al. 2018; Ohashi et al. 2018; Sadavoy et al. 2018; Takahashi et al. 2019). However, none of these show a purely radial polarization pattern throughout the entire disk, suggesting either multiple alignment mechanisms are at play (see Figure 4 for an example) or the magnetic field morphology is complex. Magnetic fields may also be detected via observations in the mid-infrared (around 10 μm; Li et al. 2016, 2017), but models describing the expected polarization morphologies at these wavelengths are lacking.

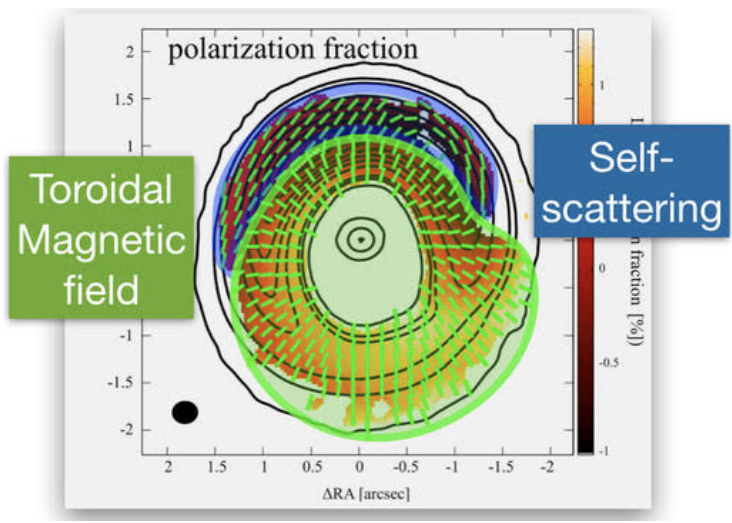

**Figure 4:** The 870 μm polarization morphology of HD142527, adapted from Ohashi et al. (2018). Models suggest that polarization in the optically thick part of the disk (highlighted in blue; Casassus et al. 2015) is likely from self-scattering while the optically thin part of the disk (highlighted in green) may be from toroidal magnetic fields (e.g., Bertrang et al. 2017).

The polarization spectrum (polarization fraction versus wavelength) for B-RAT alignment is a strong function of optical depth (e.g., Yang et al. 2017). Since optical depths of disks vary drastically for different (sub)millimeter wavelengths (e.g., Carrasco-González et al. 2016), multi-wavelength observations are essential for showing whether B-RAT alignment is taking place in these disks. Multi-wavelength observations by Alves et al. (2018) may be consistent with B-RAT alignment, but the polarization pattern in this disk is complicated and optical depth was not taken into consideration.

After combining high-resolution, multi-wavelength observations with sophisticated modeling, we will be able to confirm whether or not B-RAT alignment is occurring. Observations at long wavelengths (>2 mm) are particularly difficult, as they require very long integration times at the current interferometric facilities. Nevertheless, if B-RAT alignment is found to be an effective polarization mechanism, it would not only allow us to establish the long-sought magnetic field morphology in disks, but would also constrain the grain properties. For example, grains that are too large will not align with the magnetic field. However, if B-RAT alignment is not found, Zeeman measurements or linear polarization of spectral lines (Goldreich-Kylafis effect) may be the better avenues for observational probes of magnetic fields in disks.

**Summary and Future Needs:**
Observations have shown the existence of a wide range of disk polarization morphologies. At submillimeter wavelengths (around 870 μm), disk polarization often occurs due to self-scattering. However, observations that do not fit the simple scattering model are not easily described by any other theoretical model. Multi-wavelength, high resolution and sensitivity observations toward a large sample of disks are needed to 1) extract information about grain properties from the disk via self-scattering models; and 2) understand which other polarization mechanisms are effective. Such observations will allow us to discern how grain properties and sizes change as disks evolve toward solar systems, and could potentially unveil other important information such as magnetic field structure and dust/gas dynamics. The ability to measure both linear and circular polarization for spectral lines is also important, as this may be the most promising path for mapping and measuring magnetic fields in disks.

ALMA has the capability of measuring polarization for a large sample of disks at 1.3 mm and 870 μm (ALMA's Band 6 and 7, respectively) at around 50 au resolution. ALMA is expected to eventually implement the ability to measure polarization at shorter wavelengths (between 300 and 800 μm). However, detecting polarization at long wavelengths and at resolutions high enough to resolve disk structures is very difficult due to the lack of sensitivity. Progress on understanding polarization models would greatly benefit from upgrades to ALMA and the VLA. This includes adding more dishes, increasing continuum bandwidth, and/or adding receivers for a larger wavelength coverage.

At far-infrared wavelengths shorter than 300 μm, high-resolution observations are difficult since such observations typically require an observatory above earth's atmosphere. However, as long as there is no confusion of other sources in the field, even observations that cannot resolve a disk may help constrain grain sizes; any point in the polarimetric spectrum can help constrain grain sizes via the scattering model (e.g., Kataoka et al. 2015; 2017). SOFIA's HAWC+ polarimeter can probe polarization in isolated disks for a very select number of bright disks (perhaps less than 10). The proposed polarimeter on the Origins Space Telescope could greatly increase this number.